\documentclass[12pt]{article}
\pdfoutput=1
\usepackage[a4paper]{geometry}
\usepackage{jheppub, amsmath,amssymb,amsfonts,amsxtra, mathrsfs, makeidx,graphics,graphicx,amsthm,epsfig, ytableau,bm,longtable,float, color,tikz,mathtools,xfrac,footnote,rotating, lscape, makecell, environ,mathtools, empheq, colortbl}
\usepackage{slashed}

\usetikzlibrary{positioning}
\usetikzlibrary{chains}
\usetikzlibrary{arrows,fit,decorations.pathreplacing}
\tikzstyle{every picture}+=[remember picture]
\tikzstyle{na} = [baseline]

\usetikzlibrary{arrows, decorations.markings, calc, fadings, decorations.pathreplacing, patterns, decorations.pathmorphing, positioning}

\tikzstyle{every picture}+=[remember picture]
\tikzstyle{na} = [baseline=-.5ex]

\usepackage{bbm}
\usepackage{subfigure}

\providecommand{\abs}[1]{\lvert#1\rvert}

\numberwithin{equation}{section}

\newcommand{\be}{\begin{equation}} \newcommand{\ee}{\end{equation}}
\newcommand{\bea}{\begin{equation} \begin{aligned}} \newcommand{\eea}{\end{aligned} \end{equation}}

\def\tilde{\widetilde}

\def\a{\alpha}
\def\b{\beta}
\def\g{\gamma}

\def\ul{\underline}

\def\rt2{\sqrt{2}}

\def\abs#1{\left|#1\right|}

\def\pd{\partial}

\def\1{{\ds 1}}

\newcommand{\cF}{\mathcal{F}}

\def\O{\mathrm{O}}

\def\repa{\raise4pt\hbox{$\square$}\mkern-14mu\raise-4pt\hbox{$\square$}}
\def\repab{\overline{\raise4pt\hbox{$\square$}\mkern-14mu\raise-4pt\hbox{$\square$}\mkern-1mu}}

\def\smileface{\ensuremath{\hbox{\large$\bigcirc$}\mkern-15mu\raise-1pt\hbox{\scriptsize$\smallsmile$}%
\mkern-10mu\raise4pt\hbox{..}\mkern4mu}}
\def\frownface{\ensuremath{\hbox{\large$\bigcirc$}\mkern-15mu\raise-1pt\hbox{\scriptsize$\smallfrown$}%
\mkern-10mu\raise4pt\hbox{..}\mkern4mu}}

\newcommand{\ba}{\begin{array}}
\newcommand{\ea}{\end{array}}
\newcommand{\bi}{\begin{itemize}}
\newcommand{\ei}{\end{itemize}}

\def\bea#1\eea{\allowdisplaybreaks \begin{align}#1\end{align}}
 \newcommand{\ben}{\begin{enumerate}}
\newcommand{\een}{\end{enumerate}}
\newcommand{\bean}{\begin{eqnarray*}}
\newcommand{\eean}{\end{eqnarray*}}

\newcommand{\comment}[1]{}

\definecolor{light-gray}{gray}{0.7}

\def\aup#1 {\overset{#1}{\uparrow} \, \overset{\tilde{#1}}{\downarrow}}

\newcommand{\Dh}{\mathrm{D8}}
\newcommand{\Dz}{\mathrm{D0}}
\newcommand{\Oh}{\mathrm{O8}}

\tikzset{snake it/.style={decorate, decoration={snake, amplitude=.4mm, segment length=2mm,
                       post length=0mm,pre length=0mm}}}

\def\YM{\mathrm{Y}\mathrm{M}}

\title{Oh, wait, O8 de Sitter may be unstable!}

\author[a]{Iosif Bena,} 
\author[b]{G. Bruno De Luca,} 
\author[a]{Mariana Gra\~na}
\author[a,c]{and Gabriele Lo Monaco}
\affiliation[a]{Institut de Physique Th\'eorique, Universit\'e Paris Saclay, CEA, CNRS, \\ Orme des Merisiers, 91191 Gif-sur-Yvette CEDEX, France}
\affiliation[b]{Stanford Institute for Theoretical Physics, Stanford University,\\ Stanford, CA 94306}
\affiliation[c]{Department of Physics, Stockholm University,\\  AlbaNova, 10691 Stockholm, Sweden}
\emailAdd{iosif.bena@ipht.fr}
\emailAdd{gbdeluca@stanford.edu}
\emailAdd{mariana.grana@ipht.fr}
\emailAdd{gabriele.lomonaco@ipht.fr}

\abstract{We analyze the stability of four-dimensional de Sitter vacua constructed by compactifying massive Type IIA supergravity in the presence of two $\Oh$ sources \cite{Cordova:2018dbb}. When embedded in String Theory the first source has a clear interpretation as an $\Oh_-$ plane, but the second one could correspond to either an  $\Oh_+$ plane or to an $\Oh_-$ plane with $16\,\Dh$-branes on top. We find that this latter solution has a tachyonic instability, corresponding to the D8 branes moving away from the $\Oh_-$ plane. We comment on the possible ways of distinguishing between these sources.}

\begin{document}
\maketitle
%\newpage

%%%%%%%%%%%%%%%%%%%%%%%%%%%%%%%%%%%%%%%%%
%%%%%%%%%%%%%%%%%%%%%%%%%%%%%%%%%%%%%%%%%
%%%%%%%%%%%%%%%%%%%%%%%%%%%%%%%%%%%%%%%%%
%%%%%%%%%%%%%%%%%%%%%%%%%%%%%%%%%%%%%%%%%
\section{Introduction}

One of the most challenging open problems in String Theory is determining whether it has metastable de Sitter vacua.  An efficient way to construct such vacua is to use  an effective four-dimensional theory that incorporates String-Theory ingredients \cite{Kachru:2003aw}. 
However, the embedding of these ingredients in the full String Theory and their interactions therein are nontrivial, and have opened a rich debate about the   validity of these constructions \cite{Bena:2009xk,Bena:2011wh,Bena:2014jaa,Moritz:2017xto,Danielsson:2018ztv,Bena:2018fqc,Blumenhagen:2019qcg,Kachru:2019dvo,Bena:2019mte,Gao:2020xqh}.
In parallel to these top-down investigations, there also exist bottom-up arguments and conjectures suggesting that metastable String-Theory compactifications with a positive cosmological constant cannot be constructed \cite{Obied:2018sgi,Andriot:2018wzk,Garg:2018reu,Ooguri:2018wrx}.

A way to bypass the complications inherent to the construction of de Sitter solutions using effective four-dimensional theories and to verify the validity of these conjectures is to work directly in ten (or eleven) dimensions. Upon restricting to classical contributions to the ten-dimensional stress-energy tensor, the negative-energy sources required to evade no-go theorems \cite{Maldacena:2000mw} can be realized in String Theory as orientifold planes.  As a first approximation, one can consider the orientifold planes to be smeared along (some of) the internal directions. Even if, by definition, orientifold planes sit at fixed loci of some involution, this approximation may be justified and serve as a guide to the construction of more complete solutions \cite{Dong:2010pm,Baines:2020dmu,Marchesano:2020qvg,Junghans:2020acz}.

Solutions with explicitly localized orientifold sources are on much firmer physical ground, but constructing them is more challenging since it requires solving the field equations point-wise on the compactification manifold, and not just the integrated (averaged) version.
Furthermore, localized orientifold planes source singular supergravity solutions\footnote{See \cite[Section 2]{Cordova:2019cvf} for a recent review of orientifold singularities in supergravity.} and even when the leading-order divergence of the supergravity fields match those expected for the orientifold plane, there could still be subleading divergences that may signal deeper problems.\footnote{This happens for example with anti-D3 branes \cite{Bena:2014jaa}.}

Hence, \emph{any} supergravity solution with singularities coming from localized orientifold sources has to be understood as a good approximation of a would-be corresponding full-fledged string-theory solution only away from these singular loci. When supersymmetry is present there exist other methods to assess the validity of the supergravity approximations, but for de Sitter solutions these methods do not work.

In \cite{Cordova:2018dbb} and in \cite{Cordova:2019cvf}, C\'ordova, Tomasiello and one of the authors have constructed de Sitter backgrounds with localized O8 and O8-O6 sources respectively, which have the properties discussed above. 
We are going to refer to them as CDLT$_1$ and CDLT$_2$.

The purpose of this paper is to analyze the solutions of \cite{Cordova:2018dbb} in String Theory (CDLT$_1$), and to see whether they suffer from any instabilities. 
As we are going to review in section \ref{sec: CDLT$_1$sol}, in this class of solutions the metric takes the form
\be
\label{eq: CDLT$_1$metric}
\text{ds}^2_{\text{CDLT$_1$}} = e^{2W}\text{ds}^2_{\text{dS}_4}+e^{-2W}(\text{d}z^2+e^{2 U_2}\text{ds}^2_{M_2}+e^{2U_3}\text{ds}^2_{M_3})\,,
\ee
where the warp factors $W$ and $U_i$ (as well as the dilaton) depend on the coordinate $z$ parametrizing a circle and $M_i$ are  compact $i$-dimensional Einstein spaces; at least one between $M_2$ and $M_3$ must have negative scalar curvature.
There is a $\mathbb{Z}_2$ symmetry acting on the circle with the two fixed loci at $z=\{0, z_0\}$, where two O8 sources are located. More precisely, a source with the charges of an $\O8_+$ plane is at $z=0$ and a source with the charges of an $\O8_-$  plane is at the other fixed point (see Figure \ref{fig:circle}).\footnote{We use the standard convention where an $\O8_-$, with negative charge and tension, is the standard orientifold giving rise to an $SO(2n)$ gauge group on a stack of $n$ D8-branes on top of it, while $\O8_+$, with positive charge and tension equal to that of eight $D8$-branes, gives rise to an $Sp(2n)$ gauge group.} The existence of this de Sitter solution crucially depends on the presence of sources with the charges of these $\O8_\pm$ planes, whose negative tension violates the standard energy conditions that are incompatible with a de Sitter solution \cite{Maldacena:2000mw}.
 
As remarked above, near the orientifold sources the supergravity approximation breaks down.
However, at the source with $\O8_+$ (positive) mass and charge, the singularity is relatively mild: approaching it, all the physical quantities remain finite, with only a finite discontinuity in their derivative. Hence, this singularity is of the same type one encounters near normal D8 branes. 

This is not the situation for the O8$_-$ source, which has negative mass and charge. Very near this source the solution is both strongly coupled and strongly curved. As we will review below, the region where the supergravity approximation is not valid anymore can be made parametrically small but never made to disappear completely.
Furthermore, as one approaches the O8$_-$ plane, the subleading behavior of the supergravity fields\footnote{At higher codimension these could be divergent, much like it happens near anti-D3 brane singularities \cite{Bena:2012bk}.} deviates from the flat-space one\cite{Cordova:2018dbb, Cribiori:2019clo,Cordova:2019cvf}, and it is unclear whether they can be trusted at the supergravity level.
Besides these conceptual issues, there are also technical issues with the numerical solution near the O8$_-$ source, which we will discuss in Section \ref{sec:D8action}. Hence, our investigation will mostly steer away from the  O8$_-$ source, and focus on the $\Oh$ source with positive charge and mass, where supergravity is more trustworthy. 

We find  that at the supergravity level the non-supersymmetric CDLT$_1$ background with mobile D8-branes suffers from an instability that corresponds to the emission of a $\Dh$-brane from this $\Oh$ source. This instability is similar to the brane-jet instability \cite{Bena:2020xxb} of non-supersymmetric AdS$_4$ solutions \cite{Warner:1983du,Warner:1983vz}, except that the brane does not come out from behind a horizon but rather from a supergravity singularity.
An analogous mechanism is also responsible for the decay of an infinite class of non-supersymmetric AdS$_7$ solutions, in which case $\text{D}6$-brane sources polarize into $\Dh$-branes, destabilizing the backgrounds \cite{Apruzzi:2019ecr}.

The key question raised by our investigation is whether there is any physics that may prevent such a $\Dh$-brane from emerging from the $\Oh$ source with positive charge and mass. To address this question it is important to point out that there are two possibilities for the string-theory object corresponding to this source. 

The first possibility is that it corresponds to an $\O8_-$ plane with $16$ $\Dh$-branes on top. If this possibility is realized, the CDLT$_1$ background can be thought as just the compactified version of the very familiar configuration of two $\O8_-$ planes with $16$ $\Dh$-branes, that one obtains by T-dualizing Type-I String Theory. If so, the instability we found corresponds simply to one of the $\Dh$-branes moving away from the $\Oh$ source and this indicates that the CDLT$_1$ solution is unstable when embedded in String Theory.
\footnote{Such instabilities exist also in related models, as mentioned in \cite[Footnote 6]{Cordova:2018dbb}. Similar related models with pairs of O$p$-planes and mobile D$p$-branes have also been studied from an effective field theory point of view \cite{EvaStrings}.
Some of these models display other forms of perturbative instabilities \cite{EvaComm}.} 

The second possibility is that $\Oh$ source with positive charge and mass corresponds to an $\O8_+$ plane. If this possibility is realized the CDLT$_1$ background can be thought of as the compactified version of the $\O8_+$-$\O8_-$ configuration that one obtains by T-dualizing the Dabholkar-Park background  \cite{Dabholkar:1996pc,Witten:1997bs,Aharony:2007du}.
Furthermore, in this realization, it appears rather unlikely that the  $\O8_+$ plane could emit a D8 brane, especially because there is no String-Theory object in which it could decay, at least perturbatively.

Unfortunately supergravity is unable to distinguish between the two interpretations of the orientifold source with positive charge. Both the $\O8_+$ plane and the $\Oh_-$ plane with $16\Dh$ branes have the same charge and tension and all the  bulk supergravity fields behave identically when approaching the two singularities. These two configurations do appear distinguishable in the full String Theory, taking into account also the open string sector. It would be interesting to understand if there is any way in which this difference could manifest itself in the low-energy supergravity description. 

The paper is organized as follows. In Section \ref{sec: CDLT$_1$sol} we review the CDLT$_1$ solution. In Section \ref{sec:D8analysis} we discuss the behavior of a $\Dh$ probes in the CDLT$_1$ background: in particular, we provide the bosonic and fermionic action describing the fluctuations of a stack of $k$ probes and we comment on its stability when an $\Oh$ source is approached. We conclude in Section \ref{sec:Discussion} summarizing the results and suggesting some candidate ways to distinguish between the two possible $\Oh$ singularities.

\section{Review of CDLT$_1$ solutions}
\label{sec: CDLT$_1$sol}

In this section we review the main properties of the CDLT$_1$ solution \cite{Cordova:2018dbb} which only contain O8 sources. More details, including the full set of equations of motion and an analysis of boundary conditions can be found in \cite{Cordova:2019cvf}.

\begin{figure}[h!]
\begin{centering}
\includegraphics[scale=0.28]{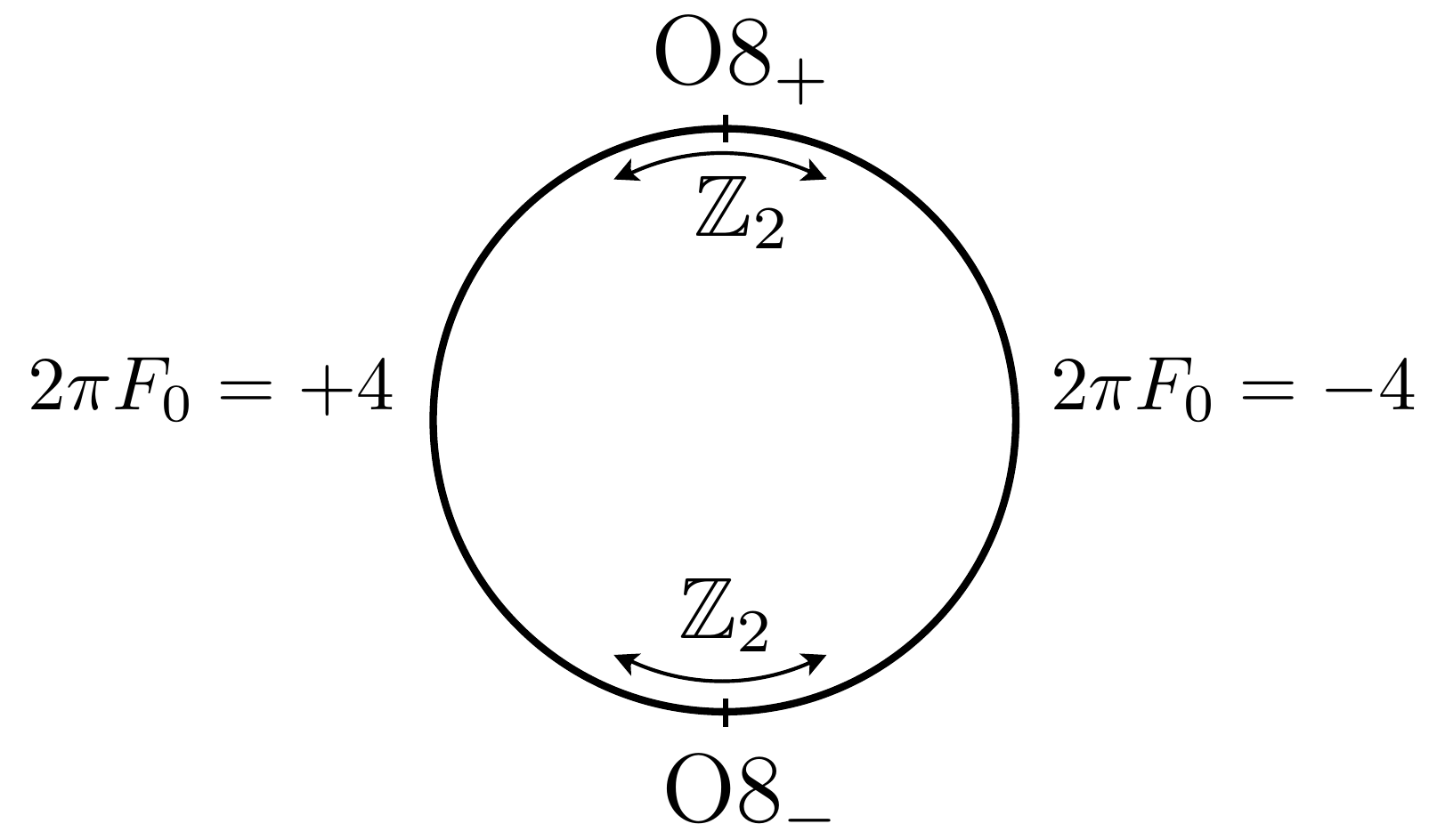}
\caption{A picture taken from \cite{Cordova:2018dbb} describing the intenrnal topology of the CDLT$_1$ solutions. Here $l_s = 1$.
}\label{fig:circle}
\end{centering}
\end{figure}

The metric of the most general CDLT$_1$ solution is:
\be
\text{ds}^2_{\text{ CDLT$_1$}} = e^{2W}\text{ds}^2_{\text{dS}_4}+e^{-2W}(\text{d}z^2+e^{2 U_2}\text{ds}^2_{M_2}+e^{2U_3}\text{ds}^2_{M_3})\,,\label{eq: metricGeneralO8}
\ee
where $M_2$($M_3$) is an Einstein manifold of dimension two(three) and $z$ is a coordinate parameterizing a circle  $z\in [0,2 z_0 ]$ with $2 z_0 \sim 0$.
All the functions depend only on the circle coordinate.
The solution has a Romans mass $F_0$ and four-form RR field strength
\begin{equation}
F_0\,=\,\frac{\pm 4}{2\pi l_s},\qquad F_4 = f_4 e^{-6W+3U_3-2U_2}dz\wedge \text{vol}_{M_3} . \label{eq:fluxesGeneralO8}
\end{equation}
Here the two values of $F_0$ correspond to the two halves of the circle (see Figure \ref{fig:circle}), on which there is a $\mathbb{Z}_2$ involution acting as the antipodal identification $z\sim -z$. 
The two fixed points at $z=0$ and $z=z_0$, are two O8-sources.
\newline
The integrated Bianchi identities require the total charge in the internal space to vanish, and thus the two orientifold sources have opposite charge. 
Similarly, the existence of a compact solution requires also the total tension to vanish, and so the charges of these  two objects are those of an $\Oh_+$ plane and an $\Oh_-$ plane.
Evaluating the equations of motion on top of the source with an O8$_+$ plane gives a constraint on the signs of the curvatures of the various Einstein factors in \eqref{eq: metricGeneralO8}.
In particular a positive four-dimensional cosmological constant (dS) is allowed only if at least one of the internal Einstein spaces has negative curvature.

There is a discrete rescaling that can be done on the solution \eqref{eq: metricGeneralO8}-\eqref{eq:fluxesGeneralO8}. 
Flux quantization requires that the integral of $F_4$ be related to an integer, $N$:
\begin{equation}
 \frac{1}{ (2\pi l_s)^3} \int_{S^1\times M_3}F_4 = N\label{eq:fluxQuantF4} .
\end{equation}
Starting from a solution where $(2\pi l_s)^{-3} \int_{S^1\times M_3}F_4 =1$, it is possible to generate other solutions with quantized flux \eqref{eq:fluxQuantF4} by 
rescaling\footnote{The rescaling constant, $c$, in %the original paper 
\cite{Cordova:2018dbb} is related to $N$ via $N=e^{4c}$.}
\begin{equation}
\label{eq:resc1}
  g_{\mu\nu} \to N^{1/2} g_{\mu\nu} ,\qquad e^{\phi} \to N^{-1/4} e^{\phi} .
\end{equation}
When the total flux, $N$, becomes large, the curvature becomes parametrically small, and the solution is weakly coupled.
This rescaling acts on the warp factors in the metric \eqref{eq: metricGeneralO8} as
\begin{equation}
\label{eq:resc2}
  e^{2W}\to N^{1/2} e^{2W},\qquad e^{2U_i}\to N e^{2 U_i},\qquad\qquad z \to N^{1/2} z
\end{equation} 
increasing the size of the circle. The solutions to the full Einstein equations were obtained numerically, but analytic expansion around the limiting points are available\footnote{One can construct analytic solutions valid everywhere as a formal expansion in $\Lambda$ \cite{Kim:2020ysx}.}.

As discussed in the Introduction, near the orientifold sources the solution is generically singular.
Near the source with positive charge the singularity is mild, similar to the one encounters when approaching D8 branes. However, since the plane with O8$_-$ charge has negative mass and negative charge, the metric and the dilaton behave as 
\be
\label{eq:asymptoteSolution}
\text{ds}^2_{\mathrm{O8}_-}\,\approx\,\frac{1}{\sqrt {r}}\left(ds^2_{\text{dS}_4}+\text{ds}^2_{M_2}+\text{ds}^2_{M_3}\right)+\sqrt{r} \text{d}r^2\,, \qquad  e^{\phi}\approx r^{-5/4},
\ee
where $r\equiv z_0 -z$. This behavior is the same as near a source with negative D8 charge and tension (such as an O8$_-$ plane) in flat space or in AdS \cite{Brandhuber:1999np,Apruzzi:2017nck,Dibitetto:2018ftj,Passias:2018zlm,Lozano:2019emq}. This can be understood as the limit $a \to 0$ of the harmonic function $e^{-4W}\equiv H = a + b\abs{r}$, which solves the Einstein equations for an O8$_-$plane with flat worldvolume. 
The discontinuity of the first derivative of this function at $r = 0$ is sourced by a $\delta$-function source.
In our more complicated solution the warping is not a simple harmonic function anymore, but the same $\delta$-function is responsible for the same asymptotic behavior near $r\sim 0$ \cite[Section 4.4]{Cordova:2019cvf}.

It is worth noticing that there exists a simple subclass obtained by tasking the two internal Einstein spaces with to have the same (negative) Einstein constant.
In these solutions the warp factors are equal $U_2=U_3\equiv U_5$ and the four-form RR field strength $F_4$ vanishes. The 5 dimensional internal space is Einstein, and the metric of the form 
\be
\label{eq: CDLT$_1$F4=0}
\text{ds}^2_{10} = e^{2W}\text{ds}^2_{\text{dS}_4}+e^{-2W}(\text{d}z^2+e^{2 U_5}\text{ds}^2_{M_5})\,. 
\ee
%%%%%%%%%%%%%%%%%%%%%%%%%%%%%%%%%%%%%%%%%
%%%%%%%%%%%%%%%%%%%%%%%%%%%%%%%%%%%%%%%%%
%%%%%%%%%%%%%%%%%%%%%%%%%%%%%%%%%%%%%%%%%
%%%%%%%%%%%%%%%%%%%%%%%%%%%%%%%%%%%%%%%%%

\section{Probing the CDLT$_1$ backgrounds with $\text{D}8$-branes}
\label{sec:D8analysis}

In this section we investigate the physics of probe D8-branes in the background  \eqref{eq: metricGeneralO8}. 

\subsection{The bosonic and fermionic action}
\label{sec:D8action}

The bosonic action of the D8 branes has a DBI and a Wess-Zumino part. From the perspective of the worldvolume D8-brane theory the interval position $z_\Dh$ of a single $\mathrm{D}8$-brane is $U(1)$ worldvolume scalar field. Besides this scalar field, the worldvolume theory has a gauge field. For $k$ $\mathrm{D}8$-branes, the gauge group of the worldvolume theory becomes $SU(k)$ and the action is\footnote{The full gauge group is  $U(k)$ but the Abelian $U(1)$ center-of-mass degrees of freedom decouple from the $SU(k)$ non-Abelian sector.}
\be
\label{eq:D8action}
S^{(B)}_{\text{D8}}\,=\,-\tau_{\text{D8}}\int d^{9}\xi\,e^{-\phi}\sqrt{-\text{det}(P[g]+\lambda\,\mathcal{F})}\pm\tau_{\text{D8}}\int P[C\wedge e^{\lambda\,\mathcal{F}}]\,,
\ee
where $\mathcal{F}$ is the worldvolume field strength and we already assumed the NSNS $B$-field to be vanishing. In the previous expression, $\tau_{\text{D8}}=2\pi/(2\pi l_s)^9$ is the brane tension while $\lambda=2\pi l_s^2$ is the worldvolume gauge coupling. Each function of $z$ can be expanded in Taylor series around a reference point, $z_0$, and each polynomial term in the series corresponds to a polynomial interaction of the worldvolume adjoint scalar field, $\Phi$, describing the positions of the D8 branes:
\be
f(z)=\sum_{i=0}^{\infty} f^{(n)}_\Dh\,(z-z_\Dh)^{n}\quad \Rightarrow\quad \sum_{i=0}^{\infty}\,f^{(n)}_\Dh\lambda^n\,\text{Tr}\Phi^n\quad \text{with}\quad f^{(n)}_\Dh=\pd_z^nf(z)|_{z_\Dh}\,.
\ee
Neglecting for the moment the contribution of $F_4$, expanding \eqref{eq:D8action} at  second order in  $\lambda$ gives:
\be
\label{eq:BosonicFinal}
\begin{split}
S^{(B)}_{\text{D8}}\,=\,-\tau_{\text{D8}}\,\lambda^2\,\text{Tr}\,\int d^{9}\xi\sqrt{-g_{(0)}}&\,\left\{ \frac{1}{\lambda^2}(\Delta_{\text{D8}}+c_9)+\frac{1}{\lambda}(\partial_z \Delta_\Dh+f_{10})\Phi\right.+\\
&\,\,\,\left.+\frac{1}{2}e^{-2W}\Delta_\Dh D_\a\Phi D^\alpha \Phi+\frac{1}{4}\Delta_\Dh  \mathcal{F}_{\alpha\beta}\mathcal{F}^{\alpha\beta}\right.+\\
&\,\,\,\left. +\frac{1}{2}(\partial^2_z\Delta_\Dh+\partial_z f_{10}) \Phi^2
\right\}+O(\lambda^3)\,,
\end{split}
\ee
where we have defined 
\be
\Delta_\Dh \equiv e^{3U_3+2U_2-W-\phi}\,,\quad F_{10} \equiv f_{10}\,\text{d}z\,\wedge\,\text{vol}_{\text{D8}}=\text{d}(c_9 \text{vol}_{\text{D8}})\,, \quad f_{10} \equiv F_0\,\Delta_\Dh\,e^{\phi-W}
\ee
and $g_{(0)}$ is the metric on $\text{AdS}_4\times M_5$ in absence of warping. Observe that worldvolume indices are raised and lowered with the worldvolume metric $g_{\alpha\beta}$ (including the warping factors) and $D_\alpha\equiv \partial_{\alpha}-i [\mathcal{A}_{\alpha}, \cdot]$.

It is convenient to use canonically normalized scalar fields, which we can obtain by defining:
\be
\tilde{\Phi}\equiv \frac{1}{g_{\YM}}\Phi\,, \quad\, g_{\YM}^2\equiv \frac{e^{-3U_3-2U_2+\phi}\big |_{z_\Dh}}{\lambda^2\tau_\Dh}\,
\ee
and to use a normalized worldvolume metric:
\be
(\gamma_\Dh)_{\alpha\beta}=\left(e^{-2W}g_{\alpha\beta}\right)\big|_{z_\Dh}\,.
\ee 
Performing the previous redefinitions and renaming for clarity $\tilde\Phi\rightarrow \Phi$, the action \eqref{eq:BosonicFinal} becomes:
\be
\label{eq:BosonicNorm}
\begin{split}
S^{(B)}_{\text{D8}}=-\text{Tr}\!\!\int\!\text{d}^{9}\xi\sqrt{-\gamma_\Dh\,}&\,\left\{-\frac{e^{4W}\,F_{\mathrm{D}8}}{\Delta_{\text{D}8}\,g_{\YM}}\Phi+\frac{1}{2}D_\a\Phi D^\alpha \Phi+\frac{1}{4\,g^2_{\YM}} \mathcal{F}_{\alpha\beta}\mathcal{F}^{\alpha\beta} -\frac{m^2_B}{2}\Phi^2\right\}\,,
\end{split}
\ee
where the worldvolume indices are now raised and lowered using $\gamma_{\Dh}$. The parameters $F_{\Dh}$  and $m_{B}$ can be identified as the force acting on the stack of brane and as the mass of the worldvolume bosonic field, and are given by:
\begin{align}
&m_{B}^2\,=\,-e^{4W}\Delta_\Dh^{-1}(\partial^2_z\Delta_\Dh+\partial_z f_{10})\\
\label{eq:ForceD8}
& F_{\mathrm{D8}}\,=\, -\lambda \tau_{\Dh}\,\Delta_\Dh(\partial_z \log\Delta_\Dh+F_0\,e^{\phi-W}) \ .
\end{align}
Let us stress that the effective Yang-Mills coupling $g_{\YM}$, the boson mass and the force are finite on the whole interval in the CDLT$_1$ backgrounds.
When the four-form is non-vanishing, one should also take into account the topological term coming from the WZ term:
\be
\delta S_\Dh^{(B)}\,=\,-\tau_{\Dh}\,\lambda^2\!\int \text{Tr}\left(C_5\wedge\cF\wedge \cF\right)\,,
\ee
where $C_5$ is the potential for the dual six-form $F_6=-\star F_4=-f_4 \text{vol}_{M_2}\wedge \text{vol}_{\text{dS}_4}$.

The fermionic action can be also computed. The details of the computation are given in Appendix \ref{sec:FermionicAction}.  The final result is:
\be
\label{eq:FermionicNorm}
\begin{split}
S_{\mathrm{D8}}^{(F)}=\frac{1}{2}\int d^9\xi\sqrt{\gamma_\Dh\,}\,\overline{\chi}\,&\left\{i\,\gamma^\alpha\nabla_\alpha\chi+\,m_F\,\chi\,+[A_\alpha,\gamma^\alpha\chi]-i\,g_{\YM}[\Phi, \chi]
\right\}\,,
\end{split}
\ee
where $\chi$ is a nine-dimensional adjoint (Majorana) spinor with mass
\begin{align}
&m_F\,=\,\frac{e^{2W}}{2}\left(\pd_z\log\Delta_\Dh+e^{\phi-W}F_0\right)\,.
\end{align}
Observe that the Wess-Zumino coupling in \eqref{eq:FermionicNorm} exactly equals the Yang-Mills coupling. In Appendix \ref{sec:FermionicAction} we also show that the fermionic action is not affected by the presence of the four-form flux.

\subsection{Approaching the singularities}

The dynamics of probe $\text{D}$-branes can unveil the presence of pathologies of a given background. Quite often, such pathologies manifest themselves as perturbative instabilities or non-perturbative decay channels  \cite{Coleman:1980aw,Maldacena:1998uz,Kachru:2002gs,Gautason:2015tla,Apruzzi:2016rny, Bena:2020xxb}.
 
\begin{figure}[h!]
\begin{centering}
\includegraphics[scale=0.54]{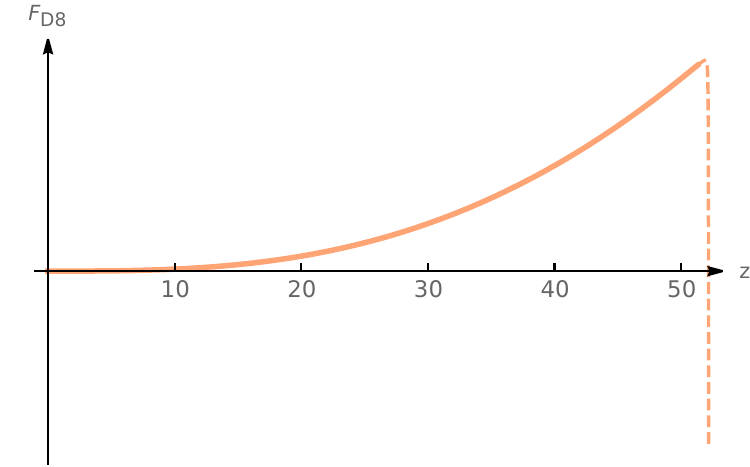}\hfill
\includegraphics[scale=0.54]{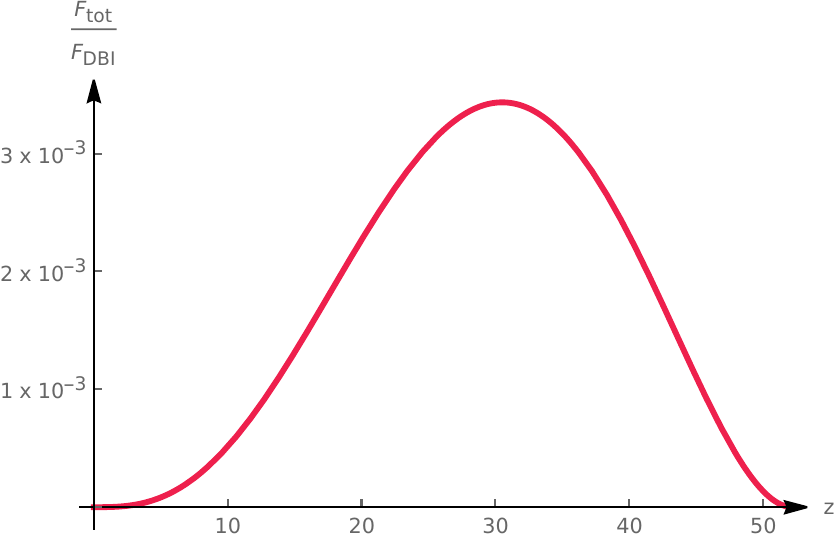}
\caption{The left figure (orange) gives the total force on a probe D8, computed as the sum of the DBI and WZ forces.
 As $z$ approaches $z_0 \sim 52$, the force drops and appears to reach a non-vanishing negative value (dashed), but we believe that this residual result is most likely a numerical artifact from the cancellation of the two diverging contributions.
 This can be seen from the right (red) figure, which gives the ratio between the total force and the gravitational attraction: $\frac{F_\textrm{tot}}{F_\textrm{DBI}}$.
 As explained in the main text, the behavior near $z \sim z_0$ has to be taken with a grain of salt, since corrections to the probe action are not negligible in the strongly-coupled region.}
\label{pic:totalForceD8s}
\end{centering}
\end{figure}

The orientifold singularities of the CDLT$_1$ solutions can be explored using $\Dh$ branes.
First, we observe that, even if the probe approaches a strongly coupled (and curved) region, the unique coupling entering the bosonic and fermionic action, $g_{\YM}$, remains finite and can be made arbitrarily small using the rescaling symmetry \eqref{eq:resc1}-\eqref{eq:resc2}, which acts as
\be
g_{\YM}\,\rightarrow\,\frac{g_{\YM}}{N^{3/2}}\,.
\ee
However, in order for the actions \eqref{eq:BosonicNorm}-\eqref{eq:FermionicNorm} to provide a meaningful effective description of the $\Dh$ worldvolume dynamics, the force $F_{\Dh}$ should be vanishing. Evaluating \eqref{eq:ForceD8} numerically for the $\text{CDLT$_1$}$ background,\footnote{The analysis is qualitatively the same also for the particular solution \eqref{eq: CDLT$_1$F4=0} where the four-form flux is vanishing.} we find  the force plotted in Figure \eqref{pic:totalForceD8s}.
\newline
\newline

\vspace*{-1cm}From  the first plot in Figure \ref{pic:totalForceD8s}, the total force appears to have a finite value at the $\Oh_-$ singularity. This would be a qualitative difference with the common behavior observed in supersymmetric solutions with $\Oh_-$ orientifolds such as \cite{Brandhuber:1999np,Apruzzi:2017nck,Dibitetto:2018ftj,Passias:2018zlm,Lozano:2019emq}.
This discrepancy arises despite the fact that the dilaton and the metric have the right behavior at that singularity. However, while a possible orientifold source can be identified by just looking at the leading components of the warp factors and the dilaton close to the singularity, the force $F_\Dh$ has a strong dependence on subleading components.
We thus believe that this result is not entirely reliable: 
First, the non-trivial profile of $F_{\Dh}$ is the result of a huge cancellation between the WZ and the DBI (electric and dilato-gravitational) forces acting on the probe, which diverge close to $z=z_0$ as 
\be
F_{\text{el, grav}}\,\propto\, \frac{1}{(z_0-z)^2}\quad \text{for}\quad z\rightarrow z_0\,
\ee
where we have used \eqref{eq:asymptoteSolution}. As a consequence, it is quite questionable how much the numerical evaluation of $F_{\Dh}$ can be trusted in the strongly curved region, where the cancellations are very large and the numerical noise is important. 

Furthermore, even if one could reach an infinite precision in evaluating the D8 action in the $\text{CDLT$_1$}$ solution, a more severe conceptual problem must be taken into account. While the divergence of the dilaton as one approaches the 
 $\Oh_-$ could naively seem not to pose a severe problem as the couplings $g_{\YM}$, $m_B^2,\, m_F$ in the effective action remain finite on the whole interval,  we have to remember that the DBI and WZ actions are only the first terms in a more general expansion in powers of $g_s \approx e^{\phi}$. In the usual supergravity regime, the higher-genus contributions can be neglected and \eqref{eq:D8action} provides a good approximation, but in a strongly coupled region such corrections cannot be considered sub-leading and become relevant. Therefore, we cannot rely on the usual effective probe action. This is exactly the situation close to the $\Oh_-$ singularity, and thus the full force $F_\Dh$ should be computed using the full genus expansion that we do not have at our disposal.
 
There is however a silver lining. In supersymmetric solutions (such as \cite{Brandhuber:1999np,Apruzzi:2017nck,Dibitetto:2018ftj,Passias:2018zlm,Lozano:2019emq}), one can invoke supersymmetry protection and obtain correct physics using the DBI+WZ approximation well beyond its validity regime. Furthermore, for anti-D3 branes in the KS solution \cite{Klebanov:2000hb}, the loop corrections to the probe potential end up canceling to all loops, giving a flat potential despite the fact that supersymmetry is broken \cite{Bena:2016fqp}. It is possible that this cancelation is related to the field content of the underlying theory, which is the same as that of the maximally supersymmetric theory \cite{Bena:2015qfa}. A similar phenomenon might make the DBI+WZ approximation of the D8 brane action describe the physics correctly even in the strongly-coupling regime.

On the other hand, we have a much better control in the region close to the orientifold source with a positive charge, $z=0$. The singularity is of exactly the (much milder) type one encounters near D8 branes: it manifests as discontinuities of the derivatives of physical quantities,  the dilaton is finite and can be made arbitrarily small if $N\gg1$.
The asymptotic behavior of the total force $F_{\Dh}$ close to $\Oh_+$ source is the following:

\be
F_{\Dh}\,=\, \alpha\,z^3+O(z^4)\, ,
\ee
with $\alpha$ a constant. The probe D8 feels no force next to the  $\Oh_+$ source, which one can attribute to the fact that in flat space they preserve the same supersymmetries. However, the D8 branes are immediately repelled if an infinitesimal displacement occurs. Hence, the probe $\Dh$-branes are actually unstable, and their worldvolume scalar parameterizing their distance away from the $\Oh_+$ source is tachyonic. 

As we explained in the Introduction, the key question is whether this probe D8 brane tachyon signals an instability of the background or is just an interesting tangential feature of the solution. This hinges on the interpretation of the orientifold source with positive D8 charge as an $\Oh_-$ plane with 16 $\Dh$ branes on top, or as an $\Oh_+$ plane.
We know that in flat un-warped space both a configuration with an $\Oh_-$ plane and an  $\Oh_+$ plane, and a configuration with two  $\Oh_-$ planes and 16 mobile D8 branes are consistent. The former is the T-dual of the Type I' Dabholkar-Park configuration \cite{Dabholkar:1996pc,Witten:1997bs,Aharony:2007du} and is dual to M-theory on the Klein bottle. The  latter is the T-dual of vanilla Type I String Theory. Furthermore, in the supergravity regime it appears impossible to distinguish between an $\Oh_+$ and an $\Oh_-$ plane with 16 $\Dh$ branes on top. 

Our analysis indicates that if the orientifold source with positive charge corresponds to an $\Oh_-$ plane with 16 $\Dh$ branes, the solution is tachyonic, and the tachyon corresponds to the D8 branes moving away from this source. Given the fact that the solution has two  $\Oh_-$ planes, it is clear that the end point of this tachyonic instability will involve a ${\mathbb Z}_2$ symmetric configuration, most likely the flat-space solution where half the D8 branes are on one of the  $\Oh_-$ planes and the other half is on the other. %\Inote{Check statement. From this configuration the D8 branes could also go back on one of the O8 planes, giving the same sources as in the dS solution} \Mnote{but then we're back to square zero. Seems like they would stay  in the middle of the two O8, but not clear to me whether there is a solution at all keeping the 5d manifold}

On the other hand, if the orientifold source with positive charge corresponds to an $\Oh_+$ plane, it does not appear possible for this plane to perturbatively emit D8 branes, and hence the instability of the D8 brane probes we found does not correspond to an instability if the solution.

\section{Discussion}
\label{sec:Discussion}

In this note we have investigated the orientifold singularities of the $\text{CDLT$_1$}$ de Sitter solutions using probe $\Dh$-branes. Starting from the usual $\text{DBI}+\text{WZ}$ action, we computed the bosonic and fermionic action describing the fluctuations of a stack of $k$ $\Dh$-branes sitting at some reference point $z_\Dh$. We observed that this action, despite its finiteness everywhere, becomes harder to trust when the stack of D8 branes approaches the singularity caused by vanishing warp factor near the $\Oh_-$ source. Accessing this region, where the action becomes strongly coupled, requires some alternative effective description that would be interesting to construct in the future. 

On the other hand, $\Dh$-branes are good probes close to the orientifold source with positive D8 charge, where the singularity is milder: in particular, the dilaton can be made arbitrarily small by increasing the units of four-form flux. The position of the $\Dh$ stack is tachyonic, and this would signal an instability of the de Sitter $\text{CDLT$_1$}$ solution if the orientifold source with positive D8 charge had the possibility to emit D8 branes. 

Since the orientifold source with positive D8 charge can correspond either to an $\Oh_+$ plane or to an $\Oh_-$ with 16 D8 branes on top, and since these two possibilities cannot be distinguished using just the supergravity solution, our work leaves open two possibilities.

\begin{itemize}
\item Either the CDLT$_1$ solutions are sourced by two $\Oh_-$ planes, one of which has $16\Dh$ branes on top. This solution is unstable.
\item Or the CDLT$_1$ solutions are sourced by an $\Oh_-$ and an $\Oh_+$ plane. This solution does not have any mobile D8 brane, and hence the probe D8 brane instability does not affect it.
 \end{itemize}

There are several ways one can try to break the tie between these possibilities. The first is to understand if there is any inconsistency upon turning on a finite worldvolume cosmological constant, in either the $\Oh_{\pm}$ configuration or in the  $2\Oh_- + 16\,\Dh$ configuration.
The second is to understand whether other probe branes can help distinguish between the two configurations. A preliminary exploration of the action of other types of probe branes in the $\text{CDLT$_1$}$ solution gives interesting physics, but unfortunately does not help us solve our conundrum. We present some of these results in Appendix B

The third is to understand whether an $\Oh_+$ plane may be related by dualities to an $\Oh_-$ plane with $16\,\Dh$ branes. These two objects have the same charge, and there exist other known situations where orientifold planes with stuck D-branes are related to other orientifold planes with the same total charge. One such example is an O3$_+$ plane, which by Seiberg-type dualities can be related to an O3$_-$ plane with a stuck D3-brane \cite{Uranga:1999ib,Elitzur:1998ju}. If such a duality relation exists between the  $\Oh_+$ and the $\Oh_-$ plane with $16\,\Dh$, the emission of a D8 brane from a $\Oh_- + 16\,\Dh$ configuration would correspond to a non-perturbative tachyonic instability of the $\Oh_-$-$\Oh_+$ configuration, which would be fascinating to figure out.

Given the importance of orientifold planes in the construction of explicit de Sitter backgrounds of String Theory, it is crucial to understand in detail the physics of these objects.
Hence another natural extension of the present work would be to investigate the richer class of CDLT$_2$ backgrounds, where the presence O6$_-$ planes avoids the introduction of singular  O8$_-$ sources.
%%%%%%%%%%%%%%%%%%%%%%%%%%%%%%%%%%%%%%%%%
%%%%%%%%%%%%%%%%%%%%%%%%%%%%%%%%%%%%%%%%%
%%%%%%%%%%%%%%%%%%%%%%%%%%%%%%%%%%%%%%%%%
%%%%%%%%%%%%%%%%%%%%%%%%%%%%%%%%%%%%%%%%%

\acknowledgments We would like to thank Emilian Duda\c s, Miguel Montero, Eva Silverstein and Alessandro Tomasiello for valuable discussions. We also thank Ivan Garozzo for useful comments on the draft.
The work of IB and MG work was partially supported by the ANR grant Black-dS-String ANR-16-CE31-0004-01, the ERC Grants 772408 ``String landscape'' and 787320 ``QBH Structure'' and the John Templeton Foundation grant 61149. The work of GBDL is supported in part by the Simons Foundation Origins of the Universe Initiative (modern inflationary cosmology collaboration) and by a Simons Investigator award. The work of GLM is supported by the Swedish Research Council grant number 2015-05333.

\appendix

\section{The Fermionic Action}
\label{sec:FermionicAction}
The fermionic action of D8 branes in flux backgrounds can be computed evaluating the action proposed in \cite{Martucci:2005rb} on the $\text{CDLT$_1$}$ backgrounds. 

Its general form is\footnote{In our conventions, $\overline\theta\,=\,i \theta^t \Gamma_{\ul 0}$.}:
\be
\label{eq:fermionS}
\begin{split}
S^{(F)}_{\text{Dp}}\,=\,\frac{\tau_{\text{Dp}}}{2}\lambda^2\int d^{p+1}\xi\,e^{-\phi}\sqrt{-\text{det}(P[g]+\mathcal{F})}&\left\{\overline{\theta}_+\left[ (M^{-1})^{\alpha\beta}\Gamma^{(P)}_\alpha \breve{\nabla}_\beta-\mathcal{E}^{(1)} \right]\theta_++\right.\\
&\,\,\,\left.-\overline{\theta}_+\,\breve{\Gamma}^{-1}_{\text{D8}}\left[(M^{-1})^{\alpha\beta}\Gamma^{(P)}_\beta \Xi_\alpha-\mathcal{E}^{(2)}\right]\theta_+\right\}\,.
\end{split}
\ee
Let us explain the meaning of the various quantities in  \eqref{eq:fermionS}: First this action is already $\kappa$-fixed, in such a way that only one chiral component of the  type IIA ten-dimensional Majorana spinor $\Theta$ supergravity is kept: in particular, we choose a $\kappa$-fixing such that $\theta_+=\frac{1}{2}(1+\Gamma_{(10)})\Theta$, with $\Gamma_{(10)}$ the ten-dimensional chirality matrix. The action is written in terms of the matrix $M_{\alpha \beta}=P[g]_{\alpha\beta}+\lambda\mathcal{F}_{\alpha \beta}$ while the pullback of the $\Gamma$-matrices is denoted by $\Gamma^{(P)}_\alpha$. 

The worldvolume coordinates are labeled by the indices $\alpha\,, \beta\,, \dots$ , while the ten-dimensional space-time coordinates are labeled by the indices  $A, B, \dots$ For a D8 brane, the other quantities in \eqref{eq:fermionS} are:
\begin{eqnarray}
&\breve\nabla_\beta=\partial_{\beta}x^{A}\,\nabla_A\,,\quad &\Xi_{\alpha}\,=\,-\frac{1}{8}e^{\phi}F_0\,\Gamma^{(P)}_\alpha\,,\\
&\mathcal{E}^{(1)}\,=\,\frac{1}{2}\Gamma^A\partial_{A}\phi\,, \quad &\mathcal{E }^{(2)}\,=\,-\frac{5}{8}e^{\phi}F_0\,.
\end{eqnarray}
If we denote the 10d vielbein with $E^{\ul A}_{A}$, the non-trivial components of the spin connection $\Omega$ are:
\be
\Omega^{\ul\alpha}{}_{\ul\beta}\,=\, \omega^{\ul\a}{}_{\ul\beta}\,,\quad \Omega^{\ul \a}{}_{\ul z}\,=\,e^{W}\pd_z (E^{\ul\alpha})\,,
\ee
where $\omega^{\ul \alpha}{}_{\ul \beta}$ is nothing but the worldvolume spin connection in the absence of warping. As a consequence, $\nabla_z = \partial_z$ % \Inote{Changed $\nabla_z\equiv \partial_z$, a consequence is not a definition}
and:
\be
\nabla_\alpha\,=\,\pd_\alpha+\frac{1}{4}\omega_{\alpha}{}^{\ul\b\,\ul \g}\Gamma_{\ul \b\,\ul \g}\,+\,\frac{1}{2}\Omega_\a{}^{\ul \b\,\ul z}\Gamma_{\ul\alpha}\Gamma_{\ul z}\,=\,\nabla_{\alpha}^{(0)}+\frac{1}{2}e^{W}\pd_{z}(\Gamma_\alpha)\Gamma_{\ul z}\,,
\ee
where ${\nabla}^{(0)}$ is the covariant derivative with $\omega$ as spin connection. At the lowest order in $\lambda$, we can use the following approximations:
\begin{eqnarray}
&M_{\alpha\,\beta}= g_{\alpha\beta}+O(\lambda)\,,\quad &\breve{\nabla}_\alpha\, = \,\nabla_{\alpha}+O(\lambda)\,,\\
&\Gamma^{(P)}_{\alpha}\, = \,\Gamma_{\alpha}+O(\lambda)\,,\quad &\Gamma_{\mathrm{D8}}^{(0)}\, = \,\Gamma_{\ul z}\Gamma_{(10)}+O(\lambda)\,.
\end{eqnarray}
We can now evaluate \eqref{eq:fermionS} for a single D8 brane to the lowest order in $\lambda$, obtaining the following Abelian action:
\be
S^{(F)}_{\mathrm{D8}}\,=\,\frac{\tau_{\mathrm{D8}}}{2}\lambda^2\,\int d^{9}\xi\sqrt{g^{(0)}}\,\Delta_\Dh\,\overline{\theta}_+\left[ \Gamma^\alpha\nabla_\alpha+\frac{1}{2}\left(e^{W}\Gamma^\alpha\pd_z\Gamma_\alpha-e^{W}\pd_z\phi\,-\,e^{\phi}F_0\right)\Gamma_{\ul z}\right]\theta_+
\ee
Following \cite{McGuirk:2012sb},  the non-Abelian action can be obtained promoting the fermion, $\theta$, to an adjoint (Majorana) fermion and performing the usual covariantization
$\pd_\alpha\rightarrow\pd_\alpha-i [A_\alpha, \cdot]+O(\lambda)$, introducing the additional coupling of the form $\Gamma^\alpha[A_\alpha, \cdot]$; consistency with $\text{T}$-duality requires an additional companion term of the form $\Gamma_z[\Phi, \cdot]$. The final result is:
\be
\begin{split}
S_{\mathrm{D8}}^{(F)}=\frac{\tau_{\mathrm{D8}}}{2}\lambda^2\int d^9\xi\sqrt{g^{(0)}}\Delta_\Dh\,\overline{\theta}_+&\left\{\Gamma^\alpha\nabla_\alpha\theta_++\frac{1}{2}\left(e^{W}\pd_z\log\Delta_\Dh+e^{\phi}F_0\right)\Gamma_{\ul z}\theta_++\right.\\
&\left.
\,{\color{white}\int}-i[A_\alpha,\Gamma^\alpha\theta_+]-i e^{-W}[\Phi, \Gamma_{\ul z}\theta_+]
\right\}
\end{split}
\ee
We can now introduce a smart choice of $\Gamma$-matrices that make manifest the $9+1$ splitting of the ten-dimensional space:
\be
\Gamma_{\ul\alpha}\,=\,\gamma_{\ul \alpha}\otimes \sigma_2\,,\quad \Gamma_{\ul z}\,=\,\mathbb{I}\otimes \sigma_1\,,
\ee
where $\gamma_{\ul \alpha}$ are 9-dimensional purely-imaginary $\gamma$-matrices with Minkowskian signature.\footnote{A way to construct them, for instance, is to start from 7-dimensional Euclidean $\gamma$-matrices, that can be chosen to be purely imaginary. Then, one can construct $\gamma_{\underline{i}}= \gamma^{(7)}_{\ul i}\otimes\sigma_3\,,$ $\gamma_{\ul 8}=\mathbb{I}\otimes \sigma_2\,,$ $\gamma_{\ul 0}=\mathbb{I}\otimes(i\sigma_1)$.} With these conventions, the 10d chirality matrix can be written as:
\be
\Gamma_{\ul 10}\,=\,\Gamma_{\ul 0}\,\dots\,\Gamma_{\ul 9}\,=\,\mathbb{I}\otimes \sigma_3\,,
\ee
where we used the fact that the 9-dimensional $\gamma$-matrices can be chosen such that $\gamma_{\ul 0}\dots\gamma_{\ul 8}=i\,\mathbb{I}$; in this basis. Hence, a chiral Majorana spinor can be decomposed as
\be
\theta_+\,=\,\chi\otimes \binom{1}{0}\,,
\ee
where $\chi$ is a 9d Majorana spinor. Using such decomposition of the spinors, it is straightforward to show that the following relations hold:
\be
\overline{\theta}\,\Gamma_{\ul\alpha}\,\theta\,=\,i\,\overline{\chi}\,\gamma_{\ul \alpha}\,\chi\,,\quad \overline{\theta}\,\Gamma_{\ul z}\,\theta\,=\,\overline \chi\,\chi\,,
\ee
where, as before, $\overline\chi=i\chi^t\gamma_{\ul 0}$ . 
As we did for the bosonic action, it is best to normalize the fermions and in particular to rescale the spinor $\chi$ by a factor $ \frac{1}{\lambda\sqrt{\tau_\Dh}}e^{(\phi-8W)/2}$. The canonically normalized action is then:
\be
\begin{split}
S_{\mathrm{D8}}^{(F)}=\frac{1}{2}\int d^9\xi\sqrt{\gamma_\Dh\,}\,\overline{\chi}\,&\left\{i\,\gamma^\alpha\nabla_\alpha\chi+\,m_F\chi\,+[A_\alpha,\gamma^\alpha\chi]-i\,g_{\YM}[\Phi, \chi]
\right\}\,.
\end{split}
\ee
where the fermion mass is defined as:
\begin{align}
&m_F\,\equiv \,\frac{e^{2W}}{2}\left(\pd_z\log\Delta_\Dh+e^{\phi-W}F_0\right)\,.
\end{align}
%\begin{figure}[h]
%\begin{centering}
 %\includegraphics[scale=0.65]{massFermion}
 %\caption{{\color{orange} $\mathbf-$} $m_{F}$}
  %\end{centering}
%\end{figure}
Finally, let us comment about a possible contribution coming from the four-form flux. In fact, the fermionic action in principle contains a Lagrangian term of the form:
\be
\delta S_{F4}\,=\, \frac{\tau_\Dh}{2}\lambda^2\int\text{d}^9\xi\,e^{-\phi}\sqrt{-\text{det}P[g]}\,\frac{e^\phi}{8}\,\overline{\theta}_+ \Gamma_{\underline{z}}[\Gamma^\alpha\slashed{F}_4\Gamma_\alpha-\slashed{F}_4]\,,
\ee
where $\slashed{F}_4\,=\,\frac{1}{4!}F_{mnpq}\Gamma^{mnpq}$. We recall that in the CDLT$_1$ backgrounds the four-form flux has the form \eqref{eq:fluxesGeneralO8}, and define for simplicity $\tilde f_4\,\equiv \,f_4 e^{-6W+3\lambda_3-2\lambda_2}$. It is easy to see that:
\be
\begin{split}
&\overline{\theta}_+\Gamma_{\underline{z}}[\Gamma^\alpha \slashed{F}_4\Gamma_\alpha-\slashed{F}_4]\theta_+\,=\,\tilde f_4\,\overline\theta_+\Gamma_{\underline z}[\Gamma^\alpha \Gamma_{\underline z}\slashed{\text{vol}}_{M_3}\Gamma_\alpha-\Gamma_{\underline z}\slashed{\text{vol}}_{M_3}]\,=\,\\
&\tilde f_4\,\overline{\theta}_+[3\slashed{\text{vol}}_{M_3}-\slashed{\text{vol}}_{M_3}]\theta_+\,=\,2\,\overline{\theta}_+\slashed{\text{vol}}_{M_3}\theta_+
\end{split}
\ee
 However, this combination vanishes for a ten-dimensional chiral spinor, given the symmetry properties of the $\Gamma$-matrices. This establishes that the non-trivial four-form flux does not affect our computation at the lowest order in $\lambda$.

  %%%%%%%%
  %%%%%%%%
  
\section{Probe $\mathrm D0$-branes}

Besides the action of probe D8 branes, it is instructive to see whether other probe D-branes have interesting physics in the CDLT$_1$ solution.
The most interesting are $\mathrm{D}0$ branes, whose action in this background naively  contains only a DBI term:  $-\tau_{\text{D}0}\int e^{-\phi}\sqrt{-g_{00}}$. However, in the presence of D8 branes (and $\Oh$ planes as well), these D0 branes come with F1 strings attached. One way to see this is from imposing  tadpole cancelation in the presence of a Romans mass \cite{Bergman:1997gf} but this does not indicate which D8 branes the F1 strings terminate on. A simpler way is to realize that in a region with no Romans mass, D0 branes have no strings attached, but passing them through $n_0$ D8 branes creates $n_0$ F1 strings via the Hanany-Witten effect \cite{Hanany:1996ie}.

\begin{figure}[h]
\begin{centering}
  \includegraphics[scale=0.6]{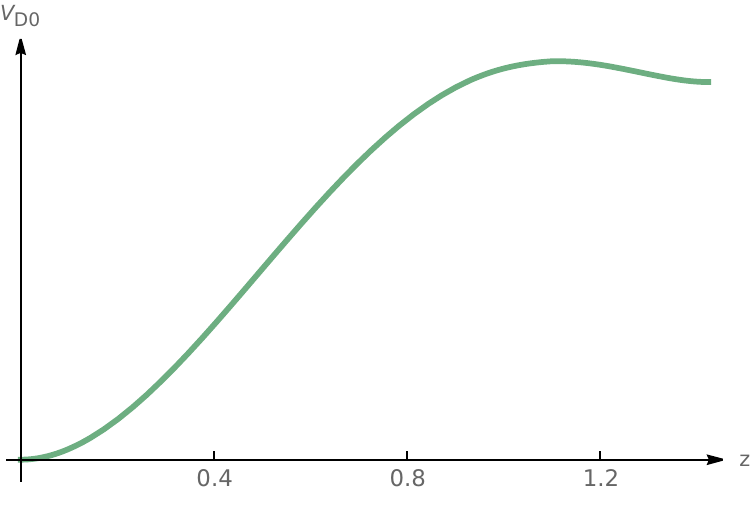}
  \definecolor{colorD0}{rgb}{0.1, 0.7, 0.}
  \caption{{\color{colorD0}$\mathbf-$} The D0-brane potential, $V_{\mathrm{D0}}$. It can be seen numerically that the unstable equilibrium point only exists when $F_4\neq 0$ and that it gets closer to $z_0$ as $N$ grows.} \label{pic:PotentialD0}
  \end{centering}
\end{figure}

Hence, the complete action describing $\text{D}0$ branes in our solution is:
\be
\label{eq:ActionD0F1}
S^{F_0}_{\text{D}0}\,=\,S_{\text{D}0}+F_0\,S_{\text{F}1}=-\tau_{\text{D}0}\int \text{d}t\,e^{-\phi}\sqrt{-g_{00}}-n_0\,\tau_{\text{F}1}\int_{\gamma} \text{d}t\text{d}\xi\,\sqrt{-\text{det}P[g]}\,,
\ee
where we already assumed that the NS potential is vanishing in the background, we denoted the temporal direction by $t$ and we defined  $n_0\equiv2\pi l_s\,F_0$. The surface $\gamma$ is such that the string extends between the $\text{D}0$ brane and a D8 brane (or an O8 plane) sourcing the Romans mass. Let us call $z_{\Dz}$ the position of the D0 brane along the interval: the tadpole cancellation requires the presence of $n_0$ attached strings ending on the orientifold plane located at $z=z_0$.
In a configuration of minimal energy, it is natural to expect that the attached string extends only along the interval transverse to the O8 planes, so that the action \eqref{eq:ActionD0F1} can be written as:
\be
\label{eq:SD0 CDLT$_1$}
S^{F_0}_{\mathrm{D}0}\,=\,-\tau_{\text{D}0}\int dt\left(e^{W(z_{\Dz})-\phi(z_{\Dz})}+F_0(z_0-z_{\Dz})\right)\,,
\ee
where we used the fact that $\tau_{\text{F}1}=\tau_{\text{D}0}/(2\pi l_s)$ and that $\text{det}P[g]=g_{00}\,g_{zz}=-1$ on the CDLT$_1$ background. The integrand in equation \eqref{eq:SD0 CDLT$_1$} can be thought as the interval-dependent potential $V_{\text{D}0}$ felt by the $\text{D}0$ branes with ($\text{F}1$) strings attached. We plot this potential in Figure \ref{pic:PotentialD0}. It is interesting to observe that the  $\mathrm{D0}$ branes are stable on both orientifold planes, and also have an unstable equilibrium point.

\bibliographystyle{ytphys}
\bibliography{ref}

\end{document}